\documentclass[aps,prl,reprint,superscriptaddress,amsmath,preprintnumbers,amssymb,floatfix]{revtex4-2}

\usepackage{graphicx}
\usepackage{physics}
\usepackage{xcolor}
\usepackage[colorlinks=true,linkcolor=blue,citecolor=blue,urlcolor=blue]{hyperref}
\usepackage[normalem]{ulem}

\newcommand{\rme}{\mathrm{e}}
\newcommand{\rmi}{\mathrm{i}}
\newcommand{\hH}{\hat{H}}
\newcommand{\hT}{\hat{T}}
\newcommand{\hN}{\hat{N}}
\newcommand{\hO}{\hat{\mathcal{O}}}
\newcommand{\tfd}{\ket{\mathrm{TFD}_\infty}}

\begin{document}

\preprint{RUP-26-32}

\title{Baby Universes Nucleate by Hiding Observers from the Boundary}

\author{Sergio E. Aguilar-Gutierrez}
\email{sergio.ernesto.aguilar@gmail.com}
\affiliation{Qubits and Spacetime Unit, Okinawa Institute of Science and Technology Graduate University, 1919-1 Tancha, Onna, Okinawa 904 0495, Japan}

\author{Takato Mori}
\email{takato.mori@yukawa.kyoto-u.ac.jp}
\affiliation{Department of Physics, Rikkyo University, 3-34-1 Nishi-Ikebukuro, Toshima-ku, Tokyo 171-8501, Japan}

\begin{abstract}
Observers are expected to be essential to describe closed universes, but their description within a conventional holographic theory remains challenging. We study two double-scaled Sachdev-Ye-Kitaev models coupled by a four-parameter interaction, which includes a generalized Maldacena-Qi coupling, in a state that contains an observer, represented as a particle. With a heavy particle inside, the wormhole connecting the two systems can only grow. Rare products of Hamiltonian strings cancel the string of the particle: a closed baby universe then carries the observer, and the wormhole, now empty, recovers discrete states of finite length. The energy gained by the empty wormhole pays for the entropy of the observer, in a first-order transition whose latent heat is the transition temperature times this entropy. Krylov complexity, which measures the wormhole length, drops at the transition and thus detects from the boundary the emergence of the baby universe, whose boundary description is actively debated. The boundary also loses the Holevo information about the microstates of the observer, which the baby universe carries.
\end{abstract}

\maketitle

\textit{Introduction.}---
How can a boundary quantum system describe a universe without one? A single state of a holographic conformal field theory (CFT) can admit two semiclassical bulk descriptions, only one of which contains a closed universe~\cite{Antonini:2024mci}, and a closed cosmology can be encoded in the entanglement between two CFTs~\cite{Antonini:2023hdh}; whether such a baby universe has a semiclassical boundary description is actively debated~\cite{Engelhardt:2025vsp,Gesteau:2025obm,Engelhardt:2026diagnostic,Higginbotham:2025dvf,Antonini:2025ioh,Liu:2025cml,Liu:2025ikq,Kudler-Flam:2025cki,Mori:2025jej,Belin:2025ako,Miyata:2026myd,Bahiru:2026gjd}. The gravitational path integral assigns a one-dimensional Hilbert space to a closed universe~\cite{Antonini:2024mci,Usatyuk:2024isz,Harlow:2025pvj,Abdalla:2025gzn}, unless it contains an observer, whose degrees of freedom then set its dimension~\cite{Harlow:2025pvj,Abdalla:2025gzn,Chen:2025fwp,Kudler-Flam:2025cki,Akers:2025observers,Higginbotham:2025clp}. Observers thus appear to be essential to describe closed universes.

\begin{figure*}[t]
    \centering
    \setlength{\tabcolsep}{4pt}
    \begin{tabular}{cccc}
        \includegraphics[height=1.4in]{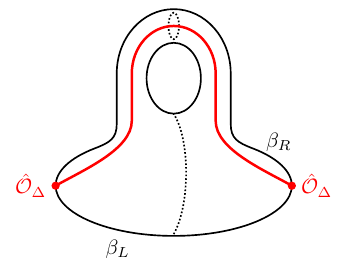} &
        \includegraphics[height=1.4in]{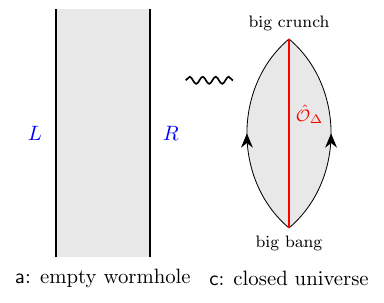} &
        \includegraphics[height=1.4in]{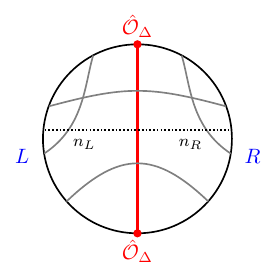} &
        \includegraphics[height=1.4in]{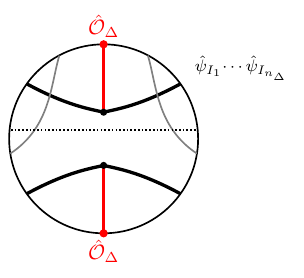}\\
        (a) & (b) & (c) & (d)
    \end{tabular}
    \caption{The baby universe saddle and its chord description. (a) Euclidean bulk: the worldline of the particle runs along a handle~\cite{Sasieta:2025vck}. (b) Lorentzian picture: an empty wormhole $\mathsf{a}$ (global AdS$_2$ in the MQ vacuum) joins the boundaries $L$ and $R$, entangled (wavy line) with a closed universe $\mathsf{c}$ that carries the particle, representing the observer. (c) Chord diagram for the norm of the HH state~\eqref{eq:HH}: the matter chord (red) joins the two insertions of the particle, and Hamiltonian chords (gray) join pairs of Hamiltonian insertions. The open chords on the time-symmetric slice (dotted) measure the length of the wormhole, which contains the particle. (d) Absorption: Hamiltonian strings (thick chords) whose product is the string of the particle, Eq.~\eqref{eq:absorption}, take the place of the matter chord, which no longer crosses the slice, as in (a) and (b).}
    \label{fig:mechanism}
\end{figure*}

Sasieta, Swingle, and Vilar L\'opez (S$^2$VL)~\cite{Sasieta:2025vck} realized a closed universe with two Sachdev-Ye-Kitaev (SYK) models~\cite{Sachdev:1992fk,kitaevTalks,Maldacena:2016upp}, dual to black holes in Jackiw-Teitelboim (JT) gravity~\cite{TEITELBOIM198341,JACKIW1985343,Mertens:2022irh}, coupled by a generalized Maldacena-Qi (MQ) interaction, whose ground state is an eternal traversable wormhole~\cite{Maldacena:2018lmt}. With a heavy particle inserted in their entangled state, a wormhole that contains the particle~\cite{Harlow:2018tqv} competes with a handle disk geometry, in which the particle runs through a closed baby universe while an empty wormhole joins the boundaries (Fig.~\ref{fig:mechanism}). The handle, suppressed by the topological factor $\rme^{-2S_0}$, with $S_0$ the extremal entropy, wins at low temperature through a first-order transition~\cite{Sasieta:2025vck}.

The double-scaling limit of SYK (DSSYK) is ultraviolet finite and solvable at all energies by chord diagrams~\cite{Berkooz:2018jqr,Berkooz:2018qkz,Lin:2022rbf,Berkooz:2024lgq}. Cutting a chord diagram on a time slice gives a state of the chord Hilbert space, on which the chord Hamiltonian, the averaged counterpart of the boundary Hamiltonian, acts. Lin identified it with the bulk Hilbert space of a two-dimensional quantum gravity, in which the number of chords is a discretized length of the wormhole between the boundaries~\cite{Lin:2022rbf}. The chord Hamiltonian and Hilbert space reproduce ones in JT gravity as $\lambda\to0$. In this framework, Sontag and Verlinde described a baby universe saddle of the original MQ model in the chord Hilbert space~\cite{Sontag:2026iiu}, and Sontag completed its chord rules at finite $\lambda$, finding a first-order transition and a big-bang/big-crunch cosmology~\cite{Sontag:2026xxg}. For a single SYK model, Bettaque \textit{et al.} described the wormhole that computes the variance of the one-point function of a heavy operator~\cite{Bettaque:2026fzu}. There, the handle arises when products of Hamiltonian strings reproduce the string of a heavy operator~\cite{Sontag:2026xxg,Bettaque:2026fzu}; see also Ref.~\cite{AALO}.

What, in the boundary theory, sets the cost of the handle, and which property of the spectrum lets it win? We answer these questions in a coupled system of SYK models in the double-scaling limit, which in general is a four parameter extension of the double-scaled S$^2$VL model, and the observer is a heavy particle and the spectrum is solvable at finite $\lambda$. The answer is a simple principle: \emph{with a heavy particle inside, the wormhole between the boundaries can only grow; once the particle is hidden in a closed universe, the wormhole, which still connects the two boundaries, settles at a finite length}. A baby universe forms when the energy gained by the empty wormhole pays for the entropy of the observer. Our main results are the following.

\textit{(i)~Spectral mechanism}: The particle removes the discrete states of the wormhole length; its absorption by Hamiltonian strings makes them available again.

\textit{(ii)~Observer entropy}: The handle is suppressed by $\rme^{-S_{\rm obs}}$, with $S_{\rm obs}$ the logarithm of the number of microstates of the observer.

\textit{(iii)~Latent heat}: The transition releases a latent heat $Q_L\simeq T_{\rm bu}S_{\rm obs}$, with $T_{\rm bu}$ its temperature.

\textit{(iv)~Information}: The boundary loses the Holevo information about the observer, which the baby universe carries.

\textit{(v)~Complexity}: Krylov complexity, a boundary measure of the wormhole length, drops at the transition, but not when the particle is trapped inside a wormhole of finite length.

\textit{(vi)~Universality}: The mechanism extends to the q-Askey deformations of DSSYK~\cite{Aguilar-Gutierrez:2026nmd}, which differ at finite $\lambda$ but share the JT limit $\lambda\to0$, where $\lambda$ plays the role of the gravitational coupling.

\noindent Details are in the End Matter and the Supplemental Material; a complete account will appear in Ref.~\cite{Aguilar-Gutierrez:inprep}.

\textit{Setup.}---
Each SYK system has $N$ Majorana fermions, $\{\hat\psi_i,\hat\psi_j\}=2\delta_{ij}$, and Hamiltonian $\hH_0=\rmi^{p/2}\sum_IJ_I\hat\psi_I$, a sum over strings $\hat\psi_I=\hat\psi_{i_1}\cdots\hat\psi_{i_p}$ with independent Gaussian couplings of variance $(J^2/\lambda)\binom{N}{p}^{-1}$, identical for the two systems, $L$ and $R$. We take $p\in4\mathbb{Z}$, so that the state $\tfd$ below is annihilated by $\hH_0^{(L)}-\hH_0^{(R)}$ and the model has the $\mathbb{Z}_4$ symmetry of the MQ model~\cite{Maldacena:2018lmt,Sasieta:2025vck}. The simplest coupling we consider is that of S$^2$VL~\cite{Sasieta:2025vck},
\begin{equation}\label{eq:model}
\begin{gathered}
    \hH_1=\hH_0^{(L)}+\hH_0^{(R)}+2\sigma_1\mathcal{E}\,\hH_{\rm MQ}~,\\
    \hH_{\rm MQ}=\rmi^{p\Delta_s}\binom{N}{p\Delta_s}^{-1}\sum_{|I|=p\Delta_s}\hat\psi^{(L)}_I\hat\psi^{(R)}_I~,
\end{gathered}
\end{equation}
with $\mathcal{E}\equiv J/\sqrt{\lambda(1-q)}$, $\lambda\equiv2p^2/N$, and $q\equiv\rme^{-\lambda}$. It couples strings of $p\Delta_s$ fermions of the two systems, with $\Delta_s=1/2$ as in S$^2$VL. Since the spectrum of DSSYK is $E=2\mathcal{E}\cos\theta$, $\theta\in[0,\pi]$~\cite{Berkooz:2018jqr,Berkooz:2024lgq}, and $\|\hH_{\rm MQ}\|\leq1$, the dimensionless $\sigma_1$ competes with $\hH_0$ at all energies. The state we consider is the Hartle-Hawking (HH) state
\begin{equation}\label{eq:HH}
    \ket{\Psi_\Delta}=\rme^{-\beta\hH_1/2}\hO_\Delta\tfd~,
\end{equation}
where $\tfd$ is the normalized maximally entangled state of the two systems, and $\hO_\Delta=\rmi^{p\Delta/2}\sum_{|I|=p\Delta}M_I\hat\psi^{(L)}_I$ inserts a particle, representing our observer, made of $p\Delta$ fermions, with independent Gaussian coefficients $M_I$ of variance $\binom{N}{p\Delta}^{-1}$; $\Delta$ is its scaling dimension.

In the double-scaling limit, $N,p\to\infty$ at fixed $\lambda$, averages over the couplings are sums over chord diagrams (Supplemental Material): each pair of Hamiltonian ($\hH_0$) insertions is joined by a (Hamiltonian) chord, the two insertions of $\hO_\Delta$ by a matter chord, and two chords of $m_1$ and $m_2$ fermions cross with weight $q^{m_1m_2/p^2}$ [Fig.~\ref{fig:mechanism}(c)]. On a time slice, the open chords form a state $\ket{\Delta;n_L,n_R}$, with $n_L$ ($n_R$) chords to the left (right) of the matter chord~\cite{Lin:2022rbf,Lin:2023trc}. Their number $n=n_L+n_R$ is the discretized length of the wormhole between the two systems~\cite{Lin:2022rbf}, a geometric meaning on which our identification of the change of topology relies. Its dynamics is generated by the chord Hamiltonian $\hT_1=\hT_1^{(L)}+\hT_1^{(R)}$, which we derive building on Ref.~\cite{Lin:2022rbf},
\begin{equation}\label{eq:chordH}
\begin{gathered}
    \hT_1^{(L)}=\frac{J}{\sqrt{\lambda}}\qty(\hat a_L+\hat a^\dagger_L+\frac{\sigma_1q^{\Delta_s(\hN+\Delta)}}{\sqrt{1-q}})~,\\
    \hat a_L=\hat\alpha_L[\hat n_L]_q+q^{\hat n_L+\Delta}\,\hat\alpha_R[\hat n_R]_q~,
\end{gathered}
\end{equation}
with $\hT_1^{(R)}$ obtained by $L\leftrightarrow R$. Here, $\hat a^\dagger_{L/R}$ adds a chord on the left/right of the matter chord, $\hat\alpha_{L/R}$ removes one, $[n]_q\equiv(1-q^n)/(1-q)$, and $\hN\equiv\hat n_L+\hat n_R$ is the total chord number operator. The first two terms are the DSSYK Hamiltonian~\cite{Lin:2022rbf}. The last one is the MQ coupling, whose chord crosses every open chord, with weight $q^{\Delta_s}$ each, and the matter chord, with $q^{\Delta_s\Delta}$: for $\sigma_1<0$, it is an attraction between the boundaries that decays exponentially with the wormhole length and is weakened by a heavy particle. As it acts on both sides, only the total $\hT_1$ generates the two-sided evolution (Supplemental Material).

\textit{Krylov chain.}---
We follow the HH state in its Krylov basis~\cite{Parker:2018yvk,Balasubramanian:2022tpr}, obtained by orthonormalizing the states $\hT_1^n\ket{K_0}$, where $\ket{K_0}\equiv\ket{\Delta;0,0}$ represents $\hO_\Delta\tfd$. In this basis, the evolution is that of a particle hopping on a half-line,
\begin{equation}\label{eq:lanczos algorithm}
    \hT_1\ket{K_n}=\mathcal{E}\qty(b_{n+1}\ket{K_{n+1}}+a_n\ket{K_n}+b_n\ket{K_{n-1}})~,
\end{equation}
with amplitudes $b_n$ and on-site potential $a_n$. Krylov, or spread, complexity measures how far the state $\ket{\psi(t)}\propto\rme^{-(\beta/2+\rmi t)\hT_1}\ket{K_0}$ has spread along the chain~\cite{Balasubramanian:2022tpr},
\begin{equation}\label{eq:spread}
    \mathcal{C}_{\rm S}(t)=\sum_nn\abs{\braket{K_n}{\psi(t)}}^2~.
\end{equation}
For the HH state, which is symmetric under $L\leftrightarrow R$, the Krylov index is the total chord number at leading order in $\lambda$, and the MQ attraction is the on-site potential, $a_n\simeq2\sigma_1q^{\Delta_s(n+\Delta)}$ (End Matter). Hence, $\lambda\mathcal{C}_{\rm S}$ is the renormalized length of the wormhole~\cite{Rabinovici:2023yex,Ambrosini:2024sre}: Krylov complexity is a boundary measure of the wormhole length. In the chord rules, the norm of the HH state is the return amplitude $Z_{\rm disk}\equiv\bra{K_0}\rme^{-\beta\hT_1}\ket{K_0}$, the disk contribution.

\begin{figure}[t]
    \centering
    \includegraphics[width=0.8\columnwidth]{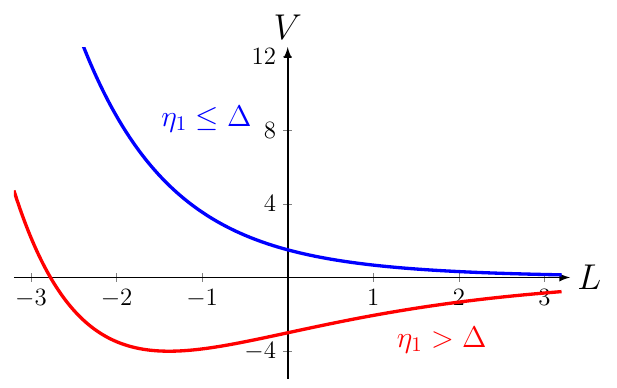}
    \caption{Morse potential~\eqref{eq:morse} of the wormhole length $L$ (not the position of the particle), which depends only on $\Delta-\eta_1$. Red: $\Delta-\eta_1=-4$, e.g., the empty wormhole at $\eta_1=4$, with four discrete states~\eqref{eq:levels}, the lowest being the MQ vacuum. Blue: $\Delta-\eta_1=1/2$, e.g., the same coupling with a particle of $\Delta=9/2$, whose repulsion fills the well and leaves only scattering states.}
    \label{fig:morse}
\end{figure}

\textit{Spectral mechanism.}---
The wormhole length behaves like a particle in a potential. In the JT regime, $\lambda\to0$ with $\eta_1\equiv-\sigma_1/\lambda$ fixed, the chord Hamiltonian near the bottom of its continuum, $E_{\rm disk}\equiv-4\mathcal{E}$, becomes $\hT_1\simeq E_{\rm disk}+2\lambda J\,[\hat k^2+V(\hat L)]$, with the renormalized length $L=\lambda n + 2\log \lambda$, 
its conjugate momentum $k$, and, for $\Delta_s=1/2$, the Morse potential of S$^2$VL~\cite{Sasieta:2025vck} (Supplemental Material)
\begin{equation}\label{eq:morse}
    V(L)=\rme^{-L}+(\Delta-\eta_1)\,\rme^{-L/2}~.
\end{equation}
The MQ interaction, $-\eta_1\rme^{-L/2}$, digs a well at short lengths (Fig.~\ref{fig:morse}), which can trap the length: below the continuum of scattering states, which describe wormholes that grow forever, there are discrete states at
\begin{equation}\label{eq:levels}
    E^{(\Delta)}_l=-\frac14\qty(\eta_1-\Delta-l-\tfrac12)^2~,\qquad 0\leq l<\eta_1-\Delta-\tfrac12~,
\end{equation}
in units of $2\lambda J$ above $E_{\rm disk}$. Without the particle, there are $l_{\rm max}+1$ of them for $\eta_1>1/2$, where $l_{\rm max}$ is explicitly given in End Matter. The lowest is the MQ vacuum, a static traversable wormhole~\cite{Maldacena:2018lmt}. We call the wormhole \emph{bound} when its length has such discrete states.

The particle repels the two boundaries, through the term $\Delta\,\rme^{-L/2}$, and makes the well shallower. The levels simply shift, $E^{(\Delta)}_l=E^{(0)}_{l+\Delta}$, also at finite $\lambda$ in the Al-Salam-Chihara (ASC) form of the chain (End Matter): the particle removes the $\Delta$ lowest discrete states of the empty wormhole. A heavy enough particle, with $\eta_1<\Delta+1/2$ as in S$^2$VL, removes all of them. The black hole, i.e., the long wormhole that contains the particle~\cite{Lin:2023trc}, then has a continuous spectrum and grows forever, whereas the spectrum of the empty wormhole starts with discrete levels~\cite{Sasieta:2025vck}. In both cases, the wormhole connects the two boundaries; only the spectrum of its length differs.

Removing the particle would thus let the length fall into the MQ vacuum, releasing its binding energy below the continuum, which we compute at finite $\lambda$ from the chain without matter (End Matter),
\begin{equation}\label{eq:Delta E}
\begin{gathered}
    \Delta E=2\mathcal{E}\qty(\abs{\rho_1}+\abs{\rho_1}^{-1}-2)~,\quad
    \rho_1\equiv\frac{\sigma_1-\sqrt{\sigma_1^2+4q^{1/2}}}{2}~.
\end{gathered}
\end{equation}
It reduces to $(\lambda J/2)(\eta_1-1/2)^2$ in the JT regime, does not depend on the particle, and is positive only if the empty wormhole is bound, $\abs{\sigma_1}>1-q^{1/2}$. However, the chord Hamiltonian~\eqref{eq:chordH} never removes the matter chord: this requires configurations that the chord rules discard.

\textit{Topology change as absorption.}---
Where is the baby universe? For each product of Hamiltonian terms, the chord rules keep only the typical configurations of fermion indices, in which different strings share a few indices at random~\cite{Berkooz:2018jqr}. Configurations in which the indices of several strings are tied to those of the particle are exponentially rare, and dropped. An exact identity keeps them: averaging over $M_I$, the norm of the HH state is
\begin{equation}\label{eq:projector}
\begin{gathered}
    Z_\Delta\equiv\overline{\braket{\Psi_\Delta}{\Psi_\Delta}}=\rme^{-S_{\rm obs}}\Tr\qty(\rme^{-\beta\hH_1}\hat P_{p\Delta})~,\quad
    \rme^{S_{\rm obs}}\equiv\binom{N}{p\Delta}~,
\end{gathered}
\end{equation}
where $\hat P_{p\Delta}\equiv\sum_{|I|=p\Delta}\hat\psi^{(L)}_I\tfd\bra{\mathrm{TFD}_\infty}\hat\psi^{(L)\dagger}_I$ projects onto the $\rme^{S_{\rm obs}}$ orthonormal states $\hat\psi^{(L)}_I\tfd$ of operator size $p\Delta$~\cite{Lin:2022rbf}. The norm is thus a thermal trace over the states created by all particles of the same size, one for each microstate of the observer.

Among the rare configurations, one dominates at low temperature. A product of strings $\hat\psi_{I_1}\cdots\hat\psi_{I_n}$ is, up to a sign, the string on the indices that appear an odd number of times, $I_1\triangle\cdots\triangle I_n$. We define $n_\Delta$ as the smallest number of Hamiltonian strings that reproduce the string $W$ of the particle,
\begin{equation}\label{eq:absorption}
    I_1\triangle\cdots\triangle I_{n_\Delta}=W~\Longleftrightarrow~\hat\psi_{I_1}\cdots\hat\psi_{I_{n_\Delta}}\hat\psi_W\propto\mathbf{1}~.
\end{equation}
For integer $\Delta$, these are $\Delta$ disjoint strings; for $p=8$ and $|W|=12$, two strings sharing two indices suffice, $n_\Delta=2$. The state $\hat\psi^{(L)}_W\tfd$ then overlaps the state $\ket{n_\Delta}$ of $n_\Delta$ open Hamiltonian chords and no matter chord, in which these chords take the place of the matter chord: they absorb it [Fig.~\ref{fig:mechanism}(d)], as in Refs.~\cite{Sontag:2026xxg,Bettaque:2026fzu}. 
The state then evolves with $\hT_{\rm MQ}\equiv\hT_1|_{\Delta=0}$, whose ground state is the MQ vacuum. Averaging also over $J_I$ (End Matter),
\begin{equation}\label{eq:two saddles}
\begin{gathered}
    Z_\Delta\simeq Z_{\rm disk}+Z_{\rm handle}~,\\
    Z_{\rm handle}\simeq\rme^{-\ell_{\rm wh}}\bra{n_\Delta}\rme^{-\beta\hT_{\rm MQ}}\ket{n_\Delta}~,
\end{gathered}
\end{equation}
where $\ket{n}$ is the normalized chord state with $n$ open Hamiltonian chords and no matter chord, and $\rme^{-\ell_{\rm wh}}\equiv\rme^{-S_{\rm obs}}P_{n_\Delta}(p\Delta)$, with $P_n(v)$ the probability that the product of the strings of $\ket{n}$ has $v$ fermions. For integer $\Delta$, on which we focus, $P_{n_\Delta}(p\Delta)=1+O(\lambda^2)$ and $\ell_{\rm wh}\simeq S_{\rm obs}$: the disk receives contributions from all the $\rme^{S_{\rm obs}}$ microstates of the observer, while the absorbed part is essentially a single state common among all the microstates.

We identify the absorbed sector with the handle of S$^2$VL [Figs.~\ref{fig:mechanism}(a) and~\ref{fig:mechanism}(b)]. The identification rests on the geometric meaning of the chord number: in both sectors, it measures the length of the wormhole between the boundaries, which, in the absorbed sector, is empty and evolves in the potential~\eqref{eq:morse} at $\Delta=0$, as in the handle saddle of S$^2$VL. Krylov complexity thus detects the change of topology from the boundary. Moreover, the suppression of the handle is set by the entropy of the observer, which plays the role of the topological $2S_0$ of JT gravity (Supplemental Material). Interference between the orderings of the strings in a chord state also imposes a selection rule (End Matter): only particles neutral under the $\mathbb{Z}_4$ symmetry, with $p\Delta$ a multiple of four, can enter the closed universe.

\begin{figure*}[t]
    \centering
    \begin{tabular}{cc}
        \includegraphics[width=0.42\textwidth]{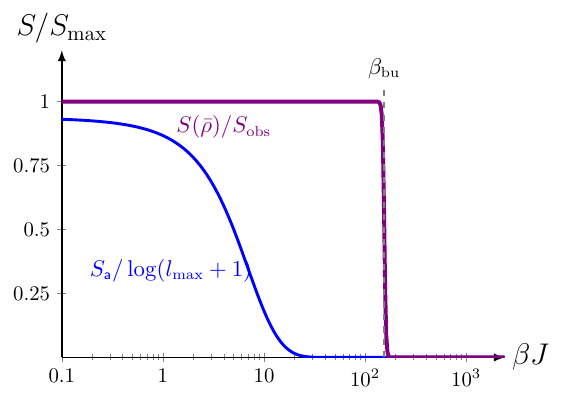} &
        \includegraphics[width=0.42\textwidth]{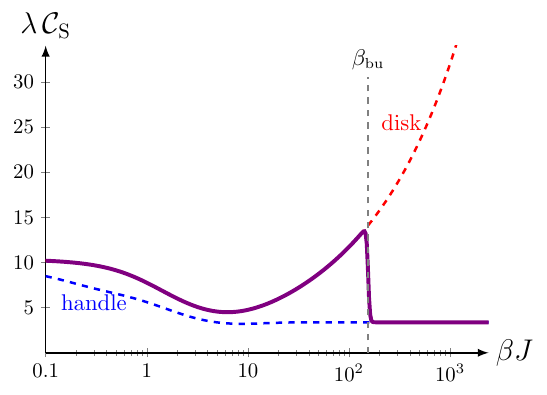}\\
        (a) & (b)
    \end{tabular}
    \caption{Boundary signatures of the transition, for $\lambda=0.3$, $\Delta_s=1/2$, $\Delta=2$, $p=8$ ($N=427$), and $\sigma_1=-0.54$, where $S_{\rm obs}=65.95$ and $\Delta E=0.324J$. The exact crossing of $Z_{\rm disk}$ and $Z_{\rm handle}$ is at $\beta_{\rm bu}J=154.5$; Eq.~\eqref{eq:beta bu} gives $154.0$ (within $0.2\%$ for $p\geq12$), and its leading term $S_{\rm obs}/\Delta E$ gives $203.5$. (a) Holevo information about the observer, $S(\bar\rho)/S_{\rm obs}$ (purple), which drops from one to zero at $\beta_{\rm bu}$, and entropy $S_{\mathsf a}$ of the empty wormhole, normalized by its maximum $\log(l_{\rm max}+1)$ over its $l_{\rm max}+1=2$ discrete states (blue), which vanishes well before. (b) Wormhole length~\eqref{eq:length} at $Jt=3$: disk sector (red), handle sector (blue), and their weighted average (purple), which drops at $\beta_{\rm bu}$ to the constant length of the MQ vacuum.}
    \label{fig:results}
\end{figure*}

\textit{First-order transition.}---
The handle wins when the energy gained by the MQ vacuum overcomes the cost $\rme^{-S_{\rm obs}}$. At low temperature, $Z_{\rm disk}\simeq\mu_{\rm disk}\rme^{-\beta E_{\rm disk}}$ is dominated by the bottom of its continuum, whose states have a weight $\mu_{\rm disk}\propto\beta^{-3/2}$, while $Z_{\rm handle}\simeq\rme^{-S_{\rm obs}}\abs{\phi_{n_\Delta}}^2\rme^{-\beta(E_{\rm disk}-\Delta E)}$ is dominated by the MQ vacuum, with $\abs{\phi_{n_\Delta}}^2$ the probability of finding $n_\Delta$ chords in it. The weight of the handle,
\begin{equation}\label{eq:xi}
    \xi\equiv\frac{Z_{\rm handle}}{Z_\Delta}\simeq\qty[1+\frac{\mu_{\rm disk}}{\abs{\phi_{n_\Delta}}^2}\rme^{S_{\rm obs}-\beta\Delta E}]^{-1}~,
\end{equation}
changes from zero to one over a window $\delta\beta\sim1/\Delta E$ around
\begin{equation}\label{eq:beta bu}
    \beta_{\rm bu}\simeq\frac{S_{\rm obs}+\log\qty(\mu_{\rm disk}/\abs{\phi_{n_\Delta}}^2)}{\Delta E}\simeq\frac{S_{\rm obs}}{\Delta E}~.
\end{equation}
By Stirling's formula, $S_{\rm obs}\simeq p\Delta\log[2\rme p/(\lambda\Delta)]$ for $p\Delta\ll N$, while the correction in Eq.~\eqref{eq:beta bu} only grows as $\log p$. The relative width $\delta\beta/\beta_{\rm bu}\sim1/S_{\rm obs}$ thus vanishes at large $p$: this is a first-order transition at $\beta_{\rm bu}J\propto p\log p$, the counterpart of the JT estimate $\beta\simeq2S_0/\abs{E_0}$ of S$^2$VL (Fig.~\ref{fig:results}). A heavier observer lowers the transition temperature.

\textit{Observer entropy and latent heat.}---
By Eq.~\eqref{eq:projector}, $Z_\Delta$ is, up to normalization, the thermal partition function of $\hH_1$ restricted to the states of size $p\Delta$. Across the transition, its mean energy $-\partial_\beta\log Z_\Delta$ drops by a latent heat $Q_L$, and its entropy $(1-\beta\partial_\beta)\log Z_\Delta$ by $\Delta S=Q_L/T_{\rm bu}$, with $T_{\rm bu}\equiv1/\beta_{\rm bu}$. In End Matter, we find
\begin{equation}\label{eq:QL DS}
    Q_L=\Delta E+O(T_{\rm bu})~,\qquad \Delta S=S_{\rm obs}+O(\log S_{\rm obs})~,
\end{equation}
so that the latent heat is the transition temperature times the entropy of the observer,
\begin{equation}\label{eq:QL}
    Q_L\simeq T_{\rm bu}S_{\rm obs}~.
\end{equation}
The heat is the binding energy of the MQ vacuum, released by the wormhole when the particle leaves. The entropy jump is a Boltzmann entropy: the $\rme^{S_{\rm obs}}$ microstates of the observer give distinct boundary states in the black hole phase, but all give the MQ vacuum in the baby universe phase. The baby universe thus stores neither heat nor thermal entropy. The transition temperature, $T_{\rm bu}\simeq\Delta E/S_{\rm obs}$, is the energy supplied by the coupling divided by the entropy supplied by the observer.

\textit{Fate of the information.}---
Can the boundaries $B=LR$ tell which microstate of the observer was prepared? Given the HH states $\ket{\Psi_I}_B=\rme^{-\beta\hH_1/2}\hat\psi^{(L)}_I\tfd_B$, one can introduce a classical register $X$ recording each microstate $I$ of the observer, the counterpart of the clone of the observer in Refs.~\cite{Harlow:2025pvj}. The full state with the external recorder is now given by a quantum--classical state 
\begin{equation}
    \sum_I p_I \dyad{\psi_I}_B\otimes \dyad{I}_X
\end{equation}
with $p_I\equiv n_I/\sum_Jn_J$, $n_I\equiv\braket{\Psi_I}{\Psi_I}$, and $\ket{\psi_I}\equiv\ket{\Psi_I}/\sqrt{n_I}$. Alternatively, this can be obtained by dephasing the purifying ancilla~\cite{Mori:2025jej}.

By discarding the microstate label $I$, we trace over $X$ and average over $M_I$ to obtain the coarse-grained state
\begin{equation}\label{eq:rho bar}
    \bar\rho_B=\sum_Ip_I\dyad{\psi_I}_B.
\end{equation}
On a disorder average, $\overline{p_I} = \rme^{-S_{\rm obs}}$. Since the $\ket{\psi_I}$ are pure, $S(\bar\rho)$ is the Holevo information of this ensemble~\cite{Holevo:1973}, i.e., the mutual information between $X$ and the boundaries $B$, which bounds the classical information about the observer’s microstate obtainable by measurements on the two boundaries~\cite{Bao:2017guc,Qi:2021oni}. As detailed in End Matter, we find
\begin{equation}\label{eq:Sobs}
    S(\bar\rho)\simeq(1-\xi)S_{\rm obs}+H(\xi)~,
\end{equation}
with $H(\xi)\equiv-\xi\log\xi-(1-\xi)\log(1-\xi)\leq\log2$: the boundaries distinguish the $\rme^{S_{\rm obs}}$ microstates of the observer in the black hole phase ($\xi\approx 0$), but not in the baby universe phase ($\xi\approx 1$), where the boundary state becomes rank-one [Fig.~\ref{fig:results}(a)] regardless of the microstate $I$ due to the absorption of the particle carrying the information.

The information that the boundary misses is the conditional entropy $S(X|B)=H(p)-S(\bar\rho)\simeq\xi S_{\rm obs}-H(\xi)$, with $H(p)\simeq S_{\rm obs}$ the Shannon entropy of $\{p_I\}_I$. In the bulk description of S$^2$VL, it is carried by the baby universe, which contains the particle: the effective dimension of the baby universe resolved by the observer, $\rme^{S(X|B)}$, grows from one to $\rme^{S_{\rm obs}}$, as found for a closed universe with an observer~\cite{Harlow:2025pvj}. 
In contrast, focusing on the handle sector ($\xi\approx 1$), we can consider a pure-state entanglement between the empty wormhole $\mathsf a$ and the baby universe $\mathsf c$ [Fig.~\ref{fig:mechanism}(b)]. The microstate dependence disappears upon normalization, thus the entanglement entropy $S_{\mathsf a}$ becomes independent of $I$ and is fixed from the energy distribution of the empty wormhole (End Matter). 
It vanishes at the transition, where the empty wormhole is in the MQ vacuum [Fig.~\ref{fig:results}(a)], as in the zero-temperature limit of S$^2$VL. 

After all, the observer entropy \(S_{\rm obs}\) plays three roles: it sets the handle suppression \(e^{-S_{\rm obs}}\), the thermodynamic entropy drop \(\Delta S\simeq S_{\rm obs}\) at the transition with the latent heat, and the amount of information about the observer’s microstate that becomes inaccessible to the boundaries and, in the bulk description, resides in the baby universe.

\textit{Complexity.}---
The time evolution does not mix the disk and handle sectors, so that the HH state is a superposition of the two, with squared norms $Z_{\rm disk}$ and $Z_{\rm handle}$. We thus define its wormhole length as
\begin{equation}\label{eq:length}
    \ell(\beta,t)\simeq(1-\xi)\,\ell_{\rm disk}(\beta,t)+\xi\,\ell_{\rm handle}(\beta,t)~,
\end{equation}
with $\ell_X\equiv\lambda\langle\hN\rangle_X\simeq\lambda\mathcal{C}_{\rm S}$ in the normalized state of sector $X$, neglecting any interference between the sectors. At $\beta=\beta_{\rm bu}$, $\ell$ drops from the linearly growing length of the long wormhole to the constant length of the MQ vacuum [Fig.~\ref{fig:results}(b)]: the particle leaves the wormhole, which shortens, as in baby universe emission in JT gravity~\cite{Saad:2019pqd}. This drop signals the change of topology as Fig.~\ref{fig:mechanism}(a).

The particle can instead be trapped inside a wormhole of finite length. Above the threshold $\abs{\rho_1}>q^{-\Delta/2}$ ($\eta_1>\Delta+1/2$), the wormhole that contains the particle also has the discrete states~\eqref{eq:levels}, and the disk sector crosses over smoothly to them at $\beta_*$, without any change of topology. Across $\beta_*$, the length at fixed time is continuous [Fig.~\ref{fig:phases}(b)], while its late-time growth turns from linear to logarithmic, and then into periodic oscillations, the closed orbits of a wormhole of finite length (End Matter). Complexity thus distinguishes the change of topology from the trapping of the particle.

\textit{Tests across couplings.}---
The absorption depends only on the combinatorics of fermion strings, not on how the two systems are coupled. The MQ term is the simplest of the q-Askey deformations of DSSYK~\cite{Aguilar-Gutierrez:2026nmd}, two-sided couplings built from $\hH_{\rm MQ}$ and the operator size (Supplemental Material). For $\Delta_s<1$, they share the JT description as $\lambda\to0$, but their spectra, and hence their transition temperatures and latent heats, differ at finite $\lambda$. With the absorption held fixed, they test its spectral consequences: a baby universe forms whenever the chain without matter has a discrete level below its continuum, which pays for the absorption. In the continuous q-ultraspherical deformation, absorption is equally allowed but gains no energy and never dominates: the spectrum decides when hiding the observer is favorable.

\textit{Discussion.}---
A closed baby universe forms when hiding the observer lets the wormhole settle into the MQ vacuum: the binding energy of the empty wormhole pays for the entropy of the observer. The combinatorics of absorption is shared with the original MQ coupling~\cite{Sontag:2026iiu,Sontag:2026xxg} and a single SYK wormhole~\cite{Bettaque:2026fzu}; our models add a solvable spectrum, in which the change of topology is traced to the discrete states of the empty wormhole, and an observer whose entropy sets the transition. Since Eq.~\eqref{eq:projector} holds at finite $N$, exact diagonalization could test the transition. A particle with nonzero $\mathbb{Z}_4$ charge cannot be absorbed: charged observers cannot enter a closed universe alone, consistent with the absence of global charges in closed universes~\cite{McNamara:2020uza,Harlow:2018tng}. Finally, the particle could serve as a clock~\cite{Sontag:2026xxg} to follow the dynamics inside the baby universe, as experienced by its observer.

\begin{acknowledgments}
We are thrilled to thank Elba Alonso-Monsalve, Micha Berkooz, Andreas Blommaert, Debarghya Chakraborty, Kanato Goto, Trivko Kukolj, Shoy Ouseph, Josef Seitz, and Jiuci Xu for inspiring discussions. The authors thank the Yukawa Institute for Theoretical Physics at Kyoto University, and the organizers of the YITP-I-25-01 workshop on ``Black Hole, Quantum Chaos and Quantum Information'' for allowing this collaboration to start. SEAG acknowledges hospitality and support from the high energy physics group at Rikkyo University during the beginning of this project. SEAG is supported by the Okinawa Institute of Science and Technology Graduate University. TM was supported by JSPS KAKENHI Grant Numbers 23KJ1154 and 24K17047. This work was made possible through the support of the WOST, WithOut SpaceTime project (\href{https://withoutspacetime.org}{https://withoutspacetime.org}), supported by Grant ID\# 63683 from the John Templeton Foundation (JTF). The opinions expressed in this work are those of the author(s) and do not necessarily reflect the views of the John Templeton Foundation. We acknowledge the use of Anthropic's Claude Sonnet 5, 5.5, Opus 5, 5.5, Fable 5, and OpenAI ChatGPT 5.6, 6, 6.1 for assistance in preparing and editing this manuscript and for numerical checks.
\end{acknowledgments}

\bibliographystyle{apsrev4-2-letter}
\bibliography{references}

\section*{End Matter}

\begin{figure*}[t]
    \centering
    \begin{tabular}{cc}
        \includegraphics[width=0.38\textwidth]{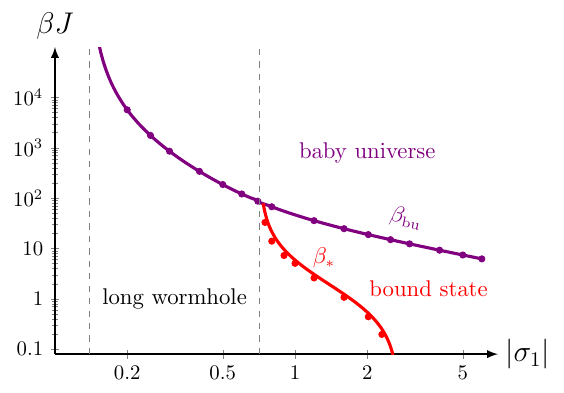} &
        \includegraphics[width=0.38\textwidth]{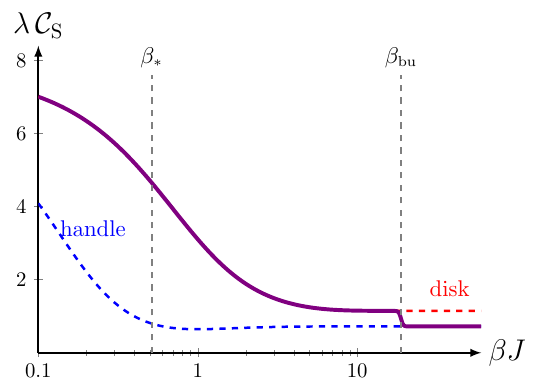}\\
        (a) & (b)
    \end{tabular}
    \caption{Regimes for $\sigma_1<0$, with $\lambda$, $\Delta_s$, $\Delta$, and $p$ as in Fig.~\ref{fig:results}. (a) Phase diagram. Purple: $\beta_{\rm bu}$ from Eq.~\eqref{eq:beta bu} (line) and from the exact crossing of $Z_{\rm disk}$ and $Z_{\rm handle}$ (markers), for $\abs{\sigma_1}>1-q^{1/2}$ (left dashed line), where the MQ vacuum is bound. Red: crossover $\beta_*$ to the discrete states of the disk sector, which exist above the threshold~\eqref{eq:threshold} (right dashed line), at leading order in $\lambda$ (line) and from the exact chain (markers), shown where it precedes $\beta_{\rm bu}$. (b) Wormhole length~\eqref{eq:length} at $Jt=3$ for $\sigma_1=-2.02$, with the colors of Fig.~\ref{fig:results}(b): it is continuous across the crossover at $\beta_*J=0.51$ and drops at the first-order transition, $\beta_{\rm bu}J=18.8$.}
    \label{fig:phases}
\end{figure*}

\textit{Krylov chain and its spectrum.}---
In the $L\leftrightarrow R$ symmetric sector, the Krylov basis of Eq.~\eqref{eq:lanczos algorithm} is, at leading order, $\ket{K_n}\propto\sum_k\binom{n}{k}\ket{\Delta;k,n-k}$~\cite{Ambrosini:2024sre}, so that the Krylov index is the total chord number. The Lanczos coefficients are
\begin{equation}\label{eq:lanczos}
    a_n\simeq2\sigma_1q^{\Delta_s(n+\Delta)}~,\quad b_n^2\simeq4\qty(1-q^{n/2})\qty(1+q^{\Delta+n/2})~,
\end{equation}
which become exact as $\lambda\to0$ at fixed $\lambda n$; at $\lambda=0.3$, $b_n^2$ deviates from its exact value by at most $2.1\%$, at $n=1$. For $\Delta_s=1/2$, this is the recurrence of the ASC polynomials~\cite{koekoek2010hypergeometric} in base $q^{1/2}$, with parameters $q^{\Delta/2}\rho_1$ and $-q^{(\Delta+1)/2}/\rho_1$, and $\rho_1$ in Eq.~\eqref{eq:Delta E}; it is exact without matter. The continuous spectrum is $E=4\mathcal{E}\cos\theta$, $\theta\in[0,\pi]$, whose lower edge is $E_{\rm disk}$. The spectrum of the ASC chord Hamiltonian with parameters $\alpha$ and $\alpha'$, $\abs{\alpha'}<1$, also has discrete states, at $E=2\mathcal{E}(x+1/x)$ for each $x=\alpha q^{l/2}$, $l=0,1,\dots$, with $\abs{x}>1$~\cite{koekoek2010hypergeometric}. Without matter, $\alpha=\rho_1$: the empty wormhole has 
$l_{\rm max}+1=\lfloor2\lambda^{-21}\log\abs{\rho_1}\rfloor+1$ discrete states for $\abs{\rho_1}>1$, i.e., $\abs{\sigma_1}>1-q^{1/2}$, and the lowest one, the MQ vacuum, gives Eq.~\eqref{eq:Delta E}. With the particle, $\alpha=q^{\Delta/2}\rho_1$, so that the discrete states are those of the empty wormhole with $l\to l+\Delta$. They exist only above the threshold
\begin{equation}\label{eq:threshold}
    \abs{\rho_1}>q^{-\Delta/2}~.
\end{equation}
In the JT regime, the ASC levels reduce to Eq.~\eqref{eq:levels}, and the threshold to $\eta_1>\Delta+1/2$ (Supplemental Material). In the window $1/2<\eta_1<\Delta+1/2$, which contains the regime of S$^2$VL, only the empty wormhole is bound: the baby universe relies on the discrete states of the sector without matter, made available by the absorption [Fig.~\ref{fig:phases}(a)].

\textit{Wormhole length.}---
Below the threshold~\eqref{eq:threshold}, the length of the long wormhole at energy $E=4\mathcal{E}\cos\theta$ grows linearly at late times, $\lambda\mathcal{C}_{\rm S}\simeq4J(\sin\theta)\,t$, as $\lambda\to0$. Above it, the discrete states lie below the energy band edge, at $E=-4\mathcal{E}\cosh\vartheta$ with $\vartheta>0$, and $\lambda\mathcal{C}_{\rm S}$ oscillates along them with period $\pi/(J\sinh\vartheta)$: these are the closed orbits of a wormhole of finite length. At the crossover $\beta_*$ between the two regimes, the growth is logarithmic, $\lambda\mathcal{C}_{\rm S}\simeq4\log(Jt)$, and at fixed time $\mathcal{C}_{\rm S}$ is continuous across it (Supplemental Material). In the MQ vacuum, the length is constant.

\textit{Absorption weights.}---
In general, $Z_{\rm handle}=\rme^{-S_{\rm obs}}\sum_nP_n(p\Delta)\bra{n}\rme^{-\beta\hT_{\rm MQ}}\ket{n}$. A chord state sums over the orderings of its strings. Two strings commute or anticommute according to the parity of their overlap, and reversing the order of the strings gives the sign $(-1)^{\sum_{a<b}\abs{I_a\cap I_b}}$, so that the orderings cancel unless the total overlap is even. Since the size of the product, $\abs{I_1\triangle\cdots\triangle I_n}$, equals $np-2\sum_{a<b}\abs{I_a\cap I_b}$ modulo four, this imposes $v\equiv np\pmod4$ on the size $v$ of the product: the chord version of the $\mathbb{Z}_4$ symmetry $\hat\psi^{(L)}_i\to-\hat\psi^{(R)}_i$, $\hat\psi^{(R)}_i\to\hat\psi^{(L)}_i$ of the MQ model~\cite{Maldacena:2018lmt,Sasieta:2025vck}. The sum over $n$ is dominated by $n=n_\Delta$: each further string requires $p/2$ more shared indices, which costs a factor of order $\rme^{-(p/2)\log p}$~\cite{Bettaque:2026fzu}. For integer $\Delta$, $n_\Delta=\Delta$ and $P_{n_\Delta}(p\Delta)=1+O(\lambda^2)$: the probability that the strings are disjoint, $\rme^{-\lambda\Delta(\Delta-1)/4}$~\cite{Bettaque:2026fzu}, is compensated at $O(\lambda)$ by the normalization of the chord state. For non-integer $\Delta$, $n_\Delta=\max(\lceil\Delta\rceil,2)$ strings must share $p(n_\Delta-\Delta)/2$ indices, an even number by the selection rule. Since two strings share only $\lambda/2$ indices on average, this costs a further factor $P_{n_\Delta}(p\Delta)$, with $-\log P_{n_\Delta}(p\Delta)$ of the same order as $S_{\rm obs}$~\cite{Aguilar-Gutierrez:inprep}. The latent heat $Q_L\simeq\Delta E$ is unchanged, but $\Delta S\simeq\ell_{\rm wh}$, so that the transition occurs at a lower temperature. Eq.~\eqref{eq:Sobs} still holds with this smaller $\xi$: the extra cost delays the transition, while the number of microstates of the observer remains $\rme^{S_{\rm obs}}$.

\textit{Transition temperature and latent heat.}---
Near the band edge, the spectral density of the disk sector vanishes quadratically in $\pi-\theta$. Hence, $\mu_{\rm disk}\propto\beta^{-3/2}$ in Eq.~\eqref{eq:xi}, and the thermal energy of the disk sector above $E_{\rm disk}$ is $3/(2\beta)$. The latent heat is then $Q_L=\Delta E+3/(2\beta_{\rm bu})+\cdots$. Since $\log Z_{\rm disk}=\log Z_{\rm handle}$ at $\beta_{\rm bu}$, $\Delta S=\beta_{\rm bu}Q_L=S_{\rm obs}+\log(\mu_{\rm disk}/\abs{\phi_{n_\Delta}}^2)+\frac32$. This gives Eq.~\eqref{eq:QL DS}, since $\log\mu_{\rm disk}=-\frac32\log\beta_{\rm bu}+\text{const.}$ and $\beta_{\rm bu}\propto S_{\rm obs}$. For Fig.~\ref{fig:results}, $\abs{\phi_2}^2=0.0028$ and $\log(\mu_{\rm disk}/\abs{\phi_2}^2)=-15.9$, so that $\Delta S=51.6$ and $Q_L=0.334J$.

\textit{Observer information.}---
In the decomposition~\eqref{eq:two saddles}, the disk components of the $\ket{\Psi_I}$ are orthogonal at leading order, with squared norm $Z_{\rm disk}$ each, while their handle components are all proportional to the same state, $c_I\rme^{-\beta\hT_{\rm MQ}/2}\ket{n_\Delta}$, with $c_I=\bra{n_\Delta}\hat\psi^{(L)}_I\tfd$ and $\sum_I\abs{c_I}^2=\rme^{S_{\rm obs}-\ell_{\rm wh}}$. The Gram matrix $\braket{\Psi_I}{\Psi_J}$ is thus a multiple of the identity plus a rank-one matrix. Writing $d=\rme^{S_{\rm obs}}$, $\bar\rho$ has one eigenvalue $a=\xi+(1-\xi)/d$ and $(d-1)$ eigenvalues $b=(1-\xi)/d$, so that
\begin{equation}\label{eq:chi finite}
    S(\bar\rho)=-a\log a-(d-1)b\log b~,
\end{equation}
which gives Eq.~\eqref{eq:Sobs} up to $O(S_{\rm obs}/d)$. The fluctuations of the overlaps between different $\ket{\Psi_I}$, suppressed by $\rme^{-S_{\rm obs}/2}$, are neglected. The entropy $S_{\mathsf a}$ is that of the energy distribution $q_E\propto\rme^{-\beta E}\abs{\braket{E}{n_\Delta}}^2$ of the empty wormhole, with $\ket{E}$ the eigenstates of $\hT_{\rm MQ}$; it is its entanglement with the baby universe if, as in S$^2$VL, the states of the latter correlated with different energies are orthogonal. Its discrete part is at most $\log(l_{\rm max}+1)$, and it vanishes once $\beta(E^{(0)}_1-E^{(0)}_0)\gg1$, i.e., already well above $T_{\rm bu}$, since $\beta_{\rm bu}(E^{(0)}_1-E^{(0)}_0)\propto S_{\rm obs}$.

\clearpage
\onecolumngrid
\setcounter{equation}{0}
\renewcommand{\theequation}{S\arabic{equation}}
\renewcommand{\theHequation}{S\arabic{equation}}
\section*{Supplemental Material}

\textit{Chord rules.}---
The average of a moment $\tr\hH_0^k$ of a single SYK system is a sum over the pairings of the $k$ insertions, drawn as chords on a circle~\cite{Berkooz:2018jqr,Berkooz:2018qkz}. Two strings of $m_1$ and $m_2$ fermions commute or anticommute according to the parity of the number of fermions they share. In the double-scaling limit, this number is Poisson distributed with mean $m_1m_2/N$, so that each crossing of their chords contributes the average sign $\rme^{-2m_1m_2/N}=q^{m_1m_2/p^2}$. The sum over diagrams is generated by the transfer matrix $\hT_0=(J/\sqrt{\lambda})(\hat a+\hat a^\dagger)$ acting on the number $n$ of open chords, with $\hat a^\dagger\ket{n}=\ket{n+1}$ and $\hat a\ket{n}=[n]_q\ket{n-1}$, so that the average of $\tr\rme^{-\beta\hH_0}$, normalized by the dimension, is $\bra{0}\rme^{-\beta\hT_0}\ket{0}$. Its spectrum is the continuum $E=2\mathcal{E}\cos\theta$, $\theta\in[0,\pi]$~\cite{Berkooz:2018jqr,Berkooz:2024lgq}. Near its edge, the dynamics of the renormalized length $\lambda n$ reduces to Liouville quantum mechanics, i.e., JT gravity~\cite{Berkooz:2018qkz,Lin:2022rbf}. A matter operator adds a chord that splits the open chords into $n_L$ and $n_R$~\cite{Lin:2022rbf,Lin:2023trc}. In Eq.~\eqref{eq:chordH}, $\hat\alpha_{L/R}$ lowers $n_{L/R}$ by one without any weight, and $\hat a_{L/R}$ sums the ways of closing a chord from the left/right boundary: closing a chord costs $q$ for each chord that it crosses and $q^\Delta$ for the matter chord. The MQ coupling acts on both systems at the same time, so that its chord crosses all the chords open at that time, which gives the last term of Eq.~\eqref{eq:chordH}. This term is shared by $\hT_1^{(L)}$ and $\hT_1^{(R)}$, which commute for decoupled systems~\cite{Lin:2022rbf}, so that
\begin{equation}\label{eq:commutator}
    [\hT_1^{(L)},\hT_1^{(R)}]=\sigma_1q^{\Delta_s\Delta}\mathcal{E}\,[\hT_0^{(L)}-\hT_0^{(R)},q^{\Delta_s\hN}]\neq0~,
\end{equation}
with $\hT_0^{(L/R)}=(J/\sqrt{\lambda})(\hat a_{L/R}+\hat a^\dagger_{L/R})$. The left and right times thus cannot be evolved independently, and the evolution of the HH state is generated by $\hT_1$.

\textit{Triple-scaling limit.}---
In the triple-scaling limit, $q^{n_{L/R}}/\lambda\equiv\rme^{-L_{L/R}}$ and $\eta_1\equiv-\sigma_1/\lambda$ are fixed as $\lambda\to0$, and the chord Hamiltonian is expanded near the lower edge $E_{\rm disk}=-4\mathcal{E}$ of its continuum. In the $L\leftrightarrow R$ symmetric sector, with $L=L_L+L_R$ and $[\hat L,\hat k]=\rmi$, this gives $\hT_1\simeq E_{\rm disk}+2\lambda J[\hat k^2+V(\hat L)]$, with the Morse potential~\eqref{eq:morse} for $\Delta_s=1/2$~\cite{Sasieta:2025vck}. Without the particle, $V$ is the potential of the gravitational mode of the eternal traversable wormhole~\cite{Maldacena:2018lmt}. Its bound states~\eqref{eq:levels} are the JT limit of the ASC levels of the End Matter. Since $\abs{\rho_1}\simeq1+(\lambda/2)(\eta_1-\tfrac12)$, the levels $E=2\mathcal{E}(x+1/x)$ with $x=\rho_1q^{l/2}$ become $E_{\rm disk}-(\lambda J/2)(\eta_1-\tfrac12-l)^2$, and those with matter follow from $\rho_1\to q^{\Delta/2}\rho_1$, i.e., $\eta_1\to\eta_1-\Delta$. The threshold~\eqref{eq:threshold} becomes $\eta_1>\Delta+1/2$, and Eq.~\eqref{eq:Delta E} becomes $\Delta E\simeq(\lambda J/2)(\eta_1-\tfrac12)^2$.

~

\textit{Length of the long wormhole.}---
Below the threshold~\eqref{eq:threshold}, the classical dynamics of the chain at energy $E=4\mathcal{E}\cos\theta$ gives, as $\lambda\to0$,
\begin{equation}\label{eq:complexity}
    \lambda\mathcal{C}_{\rm S}=2\log\frac{u+v\cosh(2J\sin\theta\,t)}{2\sin^2\theta}~,
\end{equation}
with $u\equiv1-q^\Delta-\sigma_1q^{\Delta/2}\cos\theta$ and $v^2\equiv u^2+q^\Delta(4+\sigma_1^2)\sin^2\theta$, which grows as $\lambda\mathcal{C}_{\rm S}\simeq4J\sin\theta\,t$. Above the threshold, the same dynamics at the energies $E=-4\mathcal{E}\cosh\vartheta$ of the discrete states gives closed orbits, along which $\mathcal{C}_{\rm S}$ oscillates with period $\pi/(J\sinh\vartheta)$. At the crossover between the two regimes, the growth is logarithmic, $\lambda\mathcal{C}_{\rm S}\simeq4\log(Jt)$, and at fixed time $\mathcal{C}_{\rm S}$ is continuous across it~\cite{Aguilar-Gutierrez:inprep}.

~

\textit{Observer entropy and the topological factor.}---
The number of microstates of the observer, $\binom{N}{p\Delta}$, is largest for $p\Delta=N/2$, not for the largest size $p\Delta=N$, for which it is one. There, $S_{\rm obs}=N\log2+O(\log N)=2S_\infty+O(\log N)$, with $S_\infty=(N/2)\log2$ the infinite-temperature entropy of one SYK system. The extremal entropy approaches it at large $p$, $S_0=S_\infty-\pi^2N/(4p^2)+\dots$~\cite{Maldacena:2016upp}, so that $S_0/S_\infty\to1$. Although $p\Delta=N/2$ lies beyond the double-scaling regime at fixed $\Delta$, this indicates that $\rme^{-S_{\rm obs}}$ plays the role of the topological factor $\rme^{-2S_0}$ of S$^2$VL, whose particle has $O(N)$ fermions.

~

\textit{q-Askey deformations.}---
These deformations~\cite{Aguilar-Gutierrez:2026nmd} replace the MQ term of Eq.~\eqref{eq:chordH} by functions $A_k$ and $B_k$ of $x\equiv q^{\Delta_s(\hN+\Delta)}$, with $k\leq4$ parameters $\sigma_i$,
\begin{equation}\label{eq:qAskey}
    \hT_k^{(L)}=\frac{J}{\sqrt{\lambda}}\qty(B_k(x)\,\hat a_L+\hat a^\dagger_L+\frac{A_k(x)}{\sqrt{1-q}})~.
\end{equation}
They are chosen such that, for $\Delta_s=1$ and without matter, the Krylov chain is the recurrence of q-Askey polynomials~\cite{koekoek2010hypergeometric}: ASC, continuous dual q-Hahn, and Askey-Wilson for $k=2$, $3$, and $4$. Eq.~\eqref{eq:chordH} is the case $k=1$, with $A_1=\sigma_1x$ and $B_1=1$. Their Krylov chains have $a_n\simeq2A_k$ and $b_n^2\simeq4(1-q^{n/2})(1+q^{\Delta+n/2})B_k$, at $\hN=n$ and $n-1$, respectively. For $\Delta_s<1$, they all reduce at low energies to Eq.~\eqref{eq:morse}, with $\eta_1$ replaced by the sum $\sum_i\eta_i$ of the rescaled couplings. At finite $\lambda$, their spectra depend on all the $\sigma_i$, and their chains with matter are in general not of q-Askey type. Since the absorption weights only involve the combinatorics of fermion strings, $Z_{\rm handle}$ takes the same form, with $\hT_{\rm MQ}$ replaced by Eq.~\eqref{eq:qAskey} without matter. A baby universe thus forms whenever this chain has a discrete state below its continuum; for $\Delta_s=1$, this requires $\abs{\sigma_i}>1$ for some $i$. The continuous q-ultraspherical deformation has a purely continuous spectrum, and hence no baby universe~\cite{Aguilar-Gutierrez:inprep}.

~

\textit{Information measures.}---
The register $X$ is classical, so that the Holevo information is the mutual information $I(X:B)$ of the classical-quantum state $\sum_Ip_I\ket{I}\bra{I}_X\otimes\ket{\psi_I}\bra{\psi_I}$; a coherent reference that purifies $\bar\rho$ would instead have entropy $S(\bar\rho)$. The weights $p_I$ are uniform in the black hole phase, while in the baby universe phase they approach $\abs{c_I}^2$, whose fluctuations lower $H(p)$ below $S_{\rm obs}$ by an amount of order one. Eq.~\eqref{eq:Sobs} relies on the Gram matrix of End Matter, and its accuracy for typical realizations of the couplings requires control of the fluctuations of the overlaps, beyond the averaged norms $Z_{\rm disk}$ and $Z_{\rm handle}$. The thermodynamic entropies $(1-\beta\partial_\beta)\log Z_X$ of the two sectors are defined from these normalized preparation amplitudes, and are distinct from the Holevo information and from $S_{\mathsf a}$. The interpretation of $S_{\mathsf a}$ as entanglement assumes a purification with Schmidt weights $q_E$ and orthogonal partner states in the baby universe, as in S$^2$VL~\cite{Sasieta:2025vck}. It concerns the collective length mode: the $O(N)$ entanglement of S$^2$VL just below their transition comes from light bulk fields, which chord states do not resolve~\cite{Aguilar-Gutierrez:inprep}.

~

\textit{Parameters of the figures.}---
For Figs.~\ref{fig:results} and~\ref{fig:phases}, $\lambda=0.3$, $\Delta_s=1/2$, $\Delta=2$ ($n_\Delta=2$), and $p=8$. In Fig.~\ref{fig:results}, $\sigma_1=-0.540$, so that $\rho_1=-1.236$, the disk has no discrete states, and the empty wormhole has $l_{\rm max}+1=2$ discrete states, with $E^{(0)}_1-E^{(0)}_0=0.296J$. The absorption weight gives $\ell_{\rm wh}-S_{\rm obs}=0.009$, and $\abs{\phi_2}^2=0.0028$. In Fig.~\ref{fig:phases}(b), $\sigma_1=-2.02$, so that $\rho_1=-2.39$, $l_{\rm max}+1=6$, $\Delta E=3.38J$, and $\abs{\phi_2}^2=0.26$. The disk sector is computed from the spectral density of the ASC chain, which agrees with the return amplitude on the first $4000$ sites of the Krylov chain within $10^{-5}$ in $\log Z_{\rm disk}$ for $\beta J\leq300$, and the handle sector on the first $400$ sites of the chain without matter, with the exact weights $P_n(p\Delta)$~\cite{Aguilar-Gutierrez:inprep}.

\end{document}